\documentclass[journal]{IEEEtran}

\usepackage{graphicx,amssymb,amsmath}
\usepackage{multicol}
\usepackage[noadjust]{cite}
\usepackage{setspace}
\usepackage{stfloats}
\usepackage{amsthm}
\usepackage{flushend}
\usepackage{cases,subeqnarray}
\usepackage{bm,multirow,bigstrut}
\usepackage{textcomp}
\usepackage{latexsym,bm}
\usepackage{booktabs,changebar}
\usepackage{xcolor}
\usepackage{mathtools}
\usepackage{dsfont}
\usepackage{extarrows}
\theoremstyle{plain}

\theoremstyle{plain}

\IEEEoverridecommandlockouts

\begin{document}
%----------------------------title&author&tahnks----------------------------
\title{Multiple Residual Dense Networks for Reconfigurable Intelligent Surfaces Cascaded Channel Estimation}

\author{Yu~Jin, Jiayi~Zhang, Chongwen Huang, Liang Yang, Huahua Xiao, Bo Ai, and Zhiqin Wang

%\thanks{This work was supported by the Fundamental Research Funds for the Central Universities (2019JBM016). (Corresponding Author: Jiayi Zhang, Xiaodan Zhang.) }
%\thanks{This work was supported in part by National Key R\&D Program of China under Grant 2020YFB1807201, in part by National Natural Science Foundation of China under Grants 61971027, U1834210, and 61961130391, in part by Beijing Natural Science Foundation under Grant L202013, in part by Frontiers Science Center for Smart High-speed Railway System, in part by the Royal Society Newton Advanced Fellowship under Grant NA191006, in part by the Fundamental Research Funds for the Central Universities, China, under grant number 2020JBZD005, in part by the Project of China Shenhua under grant number (GJNY-20-01-1), in part by ZTE Corporation, and State Key Laboratory of Mobile Network and Mobile Multimedia Technology. (\emph{Corresponding author: Zhiqin Wang, Jiayi Zhang.})}

%\thanks{Y. Jin and J. Zhang are with the School of Electronic and Information Engineering, Beijing Jiaotong University, Beijing 100044, China (e-mail: jiayizhang@bjtu.edu.cn).}

\thanks{Y. Jin and J. Zhang are with the School of Electronic and Information Engineering, Beijing Jiaotong University, Beijing 100044, China, and also with the Frontiers Science Center for Smart High-speed Railway System, Beijing Jiaotong University, Beijing 100044, China (e-mail: jiayizhang@bjtu.edu.cn).}% <-this % stops a space

\thanks{C. Huang is with Zhejiang Provincial Key Lab of information processing, communication and networking, Zhejiang University, Hangzhou 310007, China (e-mail: chongwenhuang@zju.edu.cn).}

\thanks{L. Yang is with College of Computer Science, and Electronic Engineering, Hunan University, Changsha 410082, China (e-mail: liangy@hnu.edu.cn).}

\thanks{H. Xiao is with ZTE Corporation, and State Key Laboratory of Mobile Network and Mobile Multimedia Technology, Shenzhen 518057, China (e-mail: xiao.huahua@zte.com.cn).}

\thanks{B. Ai is with the State Key Laboratory of Rail Traffic Control and Safety, Beijing Jiaotong University, Beijing 100044, China, and also with the Frontiers Science Center for Smart High-speed Railway System, and also with Henan Joint International Research Laboratory of Intelligent Networking and Data Analysis, Zhengzhou University, Zhengzhou 450001, China, and also with Research Center of Networks and Communications, Peng Cheng Laboratory, Shenzhen 518055, China (e-mail: boai@bjtu.edu.cn).}

\thanks{Z. Wang is with China Academy of Information and Communications Technology, Beijing 100191, P. R. China (e-mail: zhiqin.wang@caict.ac.cn).}
%}

}
\maketitle

%----------------------------abstract----------------------------
\begin{abstract}
Reconfigurable intelligent surface (RIS) constitutes an essential and promising paradigm that relies programmable wireless environment and provides capability for space-intensive communications, due to the use of low-cost massive reflecting elements over the entire surfaces of man-made structures.
However, accurate channel estimation is a fundamental technical prerequisite to achieve the huge performance gains from RIS.
By leveraging the low rank structure of RIS channels, three practical residual neural networks, named convolutional blind denoising network, convolutional denoising generative adversarial networks and multiple residual dense network, are proposed to obtain accurate channel state information,
which can reflect the impact of different methods on the estimation performance.
Simulation results reveal the evolution direction of these three methods and reveal their superior performance compared with existing benchmark schemes.

\end{abstract}

%----------------------------keywords----------------------------

%\newpage
\IEEEpeerreviewmaketitle

%----------------------------introduction----------------------------
\section{Introduction}
To greatly enhance ultra-high data rate and ubiquitous coverage requirements of the sixth-generation (6G) wireless networks, as one of the promising and innovative techniques, reconfigurable intelligent surface (RIS) aided massive multiple-input multiple-output (MIMO) is envisioned to significantly reduce link blocking probability and system energy consumption to improve link quality with sophisticated beamforming \cite{huang2019reconfigurable,wu2019towards,zhang2020prospective}.
RIS aided MIMO has been explored with near-passive array to obtain green and sustainable communications between the user equipment (UE) and the base station (BS), by appropriately and dynamically adjusting the magnitude and phase response, wireless signals can be coherently combined and steered to desired directions \cite{lin2019channel,zhang2021physical}.
Each RIS reflective element can individually control the amplitude response and phase shift of the incident electromagnetic waves at the nanosecond level to achieve energy concentration. Through reflection, refraction, absorption, and transmission, the reshaped electromagnetic waves will form new paths.
Based on the these passive and low-cost characteristics of RIS reflective elements, the RIS system requires very low energy consumption to improve the electromagnetic environment and increase propagation environment coverage.

However, benefits from a systematic performance improvement, the RIS system relies on the perfect channel state information (CSI) assumption.
Unfortunately, the above work assumes a perfect CSI but not consider the difficulty of obtaining it.
%Therefore, a low-complexity accurate channel estimation for RIS system is the fundamental key technology, especially for estimating RIS with massive passive components. Accurate CSI is essential to optimize the effective propagation of the RIS communication system.
First of all, it is quite difficult to estimate the RIS-UE and RIS-BS channels separately, unless the RIS can be equipped with radio frequency (RF) chains.
% and estimate processing can be exploited at the BS through the RIS controller.
Secondly, the cascaded channel of the RIS between the BS and the UE can be very high-dimensional due to the massive number of reflecting elements.
%partially or completely
Currently, assuming that RIS elements are connected to RF chains, the channel estimation can be performed with acceptable performance through compressed sensing (CS) based methods.
However, due to extremely low deployment, hardware and communication costs, purely passive RIS reflecting elements are undoubtedly more attractive.

By assuming active reflection patterns to achieve a smaller active array size to reduce hardware complexity, a conventional least squares method (LS) is proposed.
In addition, by using the low-rank characteristics of the MIMO channel, the training overhead can be reduced through sparse matrix decomposition. Considering the sparse representation of cascaded channels, the CS method is proposed in \cite{he2019cascaded}.
Furthermore, as the number of antennas of the UE and BS are equipped with more antennas, the channel estimation complexity increases sharply.
Using the angular-domain channel sparsity, a CS-based channel estimation scheme is proposed in \cite{chen2019channel}. However, the difference in structure sparsity between different channels will cause performance loss.
Moreover, deep learning (DL) were proposed to predict the optimal RIS phase shift matrices \cite{taha2019deep}, but it is still significant to get accurate CSI.

%recent studies have shown that it is more efficient to have a sufficient number of high-quality data sets based on learning methods.
In the field of image denoising, the previous convolutional neural network (CNN) structure can construct a pair of training images by adding synthetic noise to the noise-free images \cite{yang2018beamspace}.
%Therefore, most of the previous methods focus on classical Gaussian denoising task \cite{yang2018beamspace}.
Considering the similarity between image noise reduction and channel estimation, a deep residual learning approach was proposed to learn the cascaded channels from the noisy pilot-based observations \cite{zhang2017beyond, liu2020deep}
%image noise reduction methods have been applied in many works \cite{zhang2017beyond}.
%A deep residual learning approach was proposed to learn the cascaded channels from the noisy pilot-based observations \cite{liu2020deep}.
Also, a new architecture called Multiple Residual Dense Network (MRDN) has been proposed and has received great attention \cite{9025498}. In particular, the proposed architecture uses Residual Dense Network (RDN) as a component.

In this correspondence, we propose two practical residual neural networks to exploit the cascaded channel estimation.
Main contributions are given as follows:
First, generative adversarial networks-based convolutional blind denoising (GAN-CBD) and convolutional blind denoising network (CBDNet) are proposed to obtain accurate CSI, exploiting offline trained neural network;
%, aiming at reducing the complexity of the RIS hardware;
Second, multiple residual dense network (MRDN) is proposed to flexibly adapt to the online cascaded channel estimation;
Finally, numerical results confirm that the performance of the proposed methods can significantly outperform existing schemes in terms of ADMM and CV-DnCNN, and CBDNet.
%exploiting offline trained neural network, , which is caused by the deployment of different adjustable scales of distributed active sensing equipment exploiting one single neural network

%----------------------------system model----------------------------
\section{System Model}\label{se:model}

\begin{figure}[t]
  \centering
  %\fontsize{9}{9}\selectfont
\includegraphics[scale=0.45]{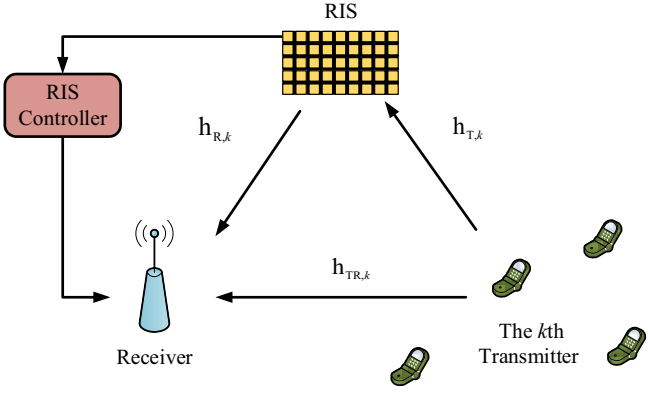}
\caption{MRDN-based channel estimation for RIS system.}
\label{fig:mrdN_ccris}
\end{figure}
We begin by considering the uplink of a time division duplex (TDD) RIS-aided mmWave communication system.
As shown in Fig. \ref{fig:mrdN_ccris}, we consider one RIS, one controller, one base station (BS) equipped with ${N_b}$ antennas and $K$ user equipments (UEs) equipped with ${N_u}$ antennas for the mmWave RIS-aided MIMO system, we assume that the planar RIS is equipped with $N = N_{\mathrm{v}} N_{\mathrm{h}}$ passive reflecting elements, where $N_{\mathrm{h}}$ and $N_{\mathrm{v}}$ denote the number of unit elements for RIS in horizontal and vertical orientations.
Defining $\mathbf{h}_{\mathrm{r}, u_k} \in \mathbb{C}^{{N} \times N_u}, \mathbf{h}_{\mathrm{r}, b} \in \mathbb{C}^{{N} \times N_b}$ as the channels from the $k$th UE to the RIS and the BS to the RIS, $\mathbf{h}_{u_k, b} \in \mathbb{C}^{{N_b} \times {N_u}}$ as the direct channel between the $k$th UE and the BS, respectively. $\mathbb{C}^{{M} \times {N}}$ represents an ${M} \times {N}$ complex-valued matrix.
Then we can express the received signal as
\begin{equation}
\mathbf{y}= \sum\limits_{k = 1}^{K} \left( \underbrace{\mathbf{h}_{\mathrm{r}, b}^T \boldsymbol{\Psi}_{k} \mathbf{h}_{\mathrm{r}, u_k} \boldsymbol{\Phi}_{k}^{\mathrm{T}}}_{\text {RIS-assisted link }}+\underbrace{\mathbf{h}_{u_k, b} \boldsymbol{\Phi}_{k}^{\mathrm{T}}}_{\text {Direct link}} \right)+ \mathbf{n},
\end{equation}
where $\mathbf{n} \sim \mathcal{CN}\left( \mathbf{0}, \sigma_n^2 \mathbf{I}\right)$ denotes the noise vector at the BS and $\sigma_n^2$ is the noise power at each antenna, $\boldsymbol{\Phi}_{k} = [\boldsymbol{\phi}_{k,1}, \boldsymbol{\phi}_{k,2}, \ldots, \boldsymbol{\phi}_{k,{N_u}}] \in \mathbb{C}^{\tau \times {N_u}}$ denotes pilot matrix for the $k$th UE, $\boldsymbol{\phi}_{n, k} \in \mathbb{C}^{\tau \times 1}$ denotes the orthogonal pilot sequence sent by the $n$th antenna of $k$th UE ($ \boldsymbol{\phi}_{k_1, i}^H \boldsymbol{\phi}_{k_2, j} = 0$, if $k_1  \neq k_2$ or $i \neq j$; $ \boldsymbol{\phi}_{k_1, i}^H \boldsymbol{\phi}_{k_2, j} = 1$, if $k_1 = k_2$ and $i = j$, $\forall k_1, k_2 \in \{1,2, \ldots, K \}$).
For transmitting the pilots, all antennas of each UE adopt different pilot sequence.
In particular, a pilot would only be allocated to one UE, resulting a orthogonal pilot matrix.
Considering a simple model where one or more users in each slot have different optimal RIS phase shift matrix. Therefore, the RIS phase shift matrix $\boldsymbol{\Psi}_{k}$ represents the phase shift introduced by the RIS to the impinging signal from the transmitter in the $k$th time slot.
%The RIS phase shift matrix $\boldsymbol{\Psi}_{k}$ represents the phase shift introduced by the RIS to the impinging signal from the transmitter in the $k$th time slot.
In addition, $\boldsymbol{\Psi}_{k} \triangleq \operatorname{diag}\{\boldsymbol{\psi}_{k}\} \in \mathbb{C}^{N \times N}$, with $\boldsymbol{\psi}_{k} \in \mathbb{C}^{N \times 1}$ representing the effective phase shifts of the RIS reflecting elements and its $n$th element is $[\boldsymbol{\psi}_{k}]_{n}=\varpi_{n} e^{j \theta_{n}}, \forall n \in \{1,2, \ldots, N \}$.
Without loss of generality, we can assume $\boldsymbol{\psi}_k = \mathbf{1}$.

By exploiting $\boldsymbol{\Phi}_{k}^{\mathrm{T}} \boldsymbol{\Phi}_{k}^{*} = \mathbf{I}$ and for simplifying the designing and analysing of the channel estimation algorithms in this work, we assume that there is no direct link between UE and BS due to blockages or negligible receive power, then, the processed received signal of the $k$th UE at the BS is given by
\begin{equation}
\begin{aligned}
\mathbf{y}_{k} = \mathbf{y} \boldsymbol{\Phi}_{k}^{*} = \mathbf{h}_{\mathrm{r}, b}^T \boldsymbol{\Psi}_{k} \mathbf{h}_{\mathrm{r}, u} + \mathbf{n}  \boldsymbol{\Phi}_{k}^{*}.
\end{aligned}
\end{equation}
Since practical mmWave channels usually have limited number of scatters, a LoS is expected in RIS systems. The mmWave channel of the $k$th UE to the RIS and BS to the RIS are, respectively, given as
 \begin{equation}
     {\mathbf{h}_{\mathrm{r}, u}} = \sum\limits_{l = 1}^{L_\mathrm{T}} {z_l \boldsymbol{\alpha}_{\mathrm{R},t} \left( {{\phi_{\mathrm{R},l}^{\mathrm{azi}}},{\phi_{\mathrm{R},l}^{\mathrm{ele}}}} \right)\boldsymbol{\alpha}_{\mathrm{T},t}^H \left( {{\phi_{\mathrm{T},l}^{\mathrm{azi}}},{\phi_{\mathrm{T},l}^{\mathrm{ele}}}} \right)},
 \end{equation}
 \begin{equation}
     {\mathbf{h}_{\mathrm{r}, b}} = \sum\limits_{l = 1}^{L_\mathrm{R}} {z_l \boldsymbol{\alpha}_{\mathrm{R},r} \left( {{\phi_{\mathrm{R},l}^{\mathrm{azi}}},{\phi_{\mathrm{R},l}^{\mathrm{ele}}}} \right)\boldsymbol{\alpha}_{\mathrm{T},r}^H \left( {{\phi_{\mathrm{T},l}^{\mathrm{azi}}},{\phi_{\mathrm{T},l}^{\mathrm{ele}}}} \right)},
 \end{equation}
where ${L} \ll \min \left( {N_{\text{act}},N_t} \right)$ denotes the number of multipaths, $z_{l, k} \in \mathbb{C}$ denotes the distance-dependent pathloss of the $\mathbf{h}_{\mathrm{T}, k}$ in the $l$th path.
$\phi_{\mathrm{R},l}^{\mathrm{ele}}$ $(\phi_{\mathrm{R},l}^{\mathrm{azi}})$ denotes the elevation (azimuth) angle-of-arrival of the $l$th path for both $\mathbf{h}_{\mathrm{T}, k}$ and $\mathbf{h}_{\mathrm{T}, k_{\text{act}} }$.
The steering vectors depend on the array geometry. For the typical channel ${\mathbf{h}_{\mathrm{T}, k_{\text{act}}}}$ and ${\mathbf{h}_{\mathrm{T}, k}}$, variables $\boldsymbol{\alpha}_{\mathrm{R}} ( {{\phi_{\mathrm{R},l}^{\mathrm{azi}}}, {\phi_{\mathrm{R},l}^{\mathrm{ele}}}} ) \in \mathbb{C}^{{N_r} \times 1}$ and $\boldsymbol{\alpha}_{\mathrm{T}} ( {{\phi_{\mathrm{T},l}^{\mathrm{azi}}},{\phi_{\mathrm{T},l}^{\mathrm{ele}}}} ) \in \mathbb{C}^{{N_t} \times 1}$ are given by
\begin{equation}
\begin{aligned}
&\boldsymbol{\alpha}_{\mathrm{R}, t}\left(\phi_{\mathrm{R}, l}^{\mathrm{azi}}, \phi_{\mathrm{R}, l}^{\mathrm{ele}}\right)= [1, \mathrm{e}^{j 2 \pi k d \sin \phi_{\mathrm{R}, l}^{\text {azi }} \sin \phi_{\mathrm{R}, l}^{\text {ele }} / \lambda}, \cdots, \\
& \mathrm{e}^{j 2 \pi d\left( N_\mathrm{v} -1   \right)   \sin \phi_{\mathrm{R}, l}^{\text {axi }} \! \sin \phi_{\mathrm{R}, l}^{\text {ele }} / \!  \lambda}]^{T}  \!  \! \otimes \!  [1, \mathrm{e}^{j 2 \pi k d \cos \phi_{\mathrm{R}, l}^{\text {ele }} / \lambda}, \! \cdots \! , \\
& \mathrm{e}^{j 2 \pi d\left(N_\mathrm{h} -1\right) \cos \phi_{\mathrm{R}, l}^{\text {ele }} / \lambda}]^{T},
\end{aligned}
 \end{equation}
\begin{equation}
\begin{aligned}
& \boldsymbol{\alpha}_{\mathrm{T}, t}\left(\phi_{\mathrm{T}, l}^{\mathrm{azi}}, \phi_{\mathrm{T}, l}^{\mathrm{ele}}\right)=[1, \mathrm{e}^{j 2 \pi d \sin \phi_{\mathrm{T}, l}^{\mathrm{azi}} \sin \phi_{\mathrm{T}, l}^{\mathrm{ele}} / \lambda}, \cdots, \\
& \mathrm{e}^{j 2 \pi d\left(N_{ T1}-1\right) \sin \phi_{\mathrm{T}, l}^{\mathrm{ai}} \sin \phi_{\mathrm{T}, l}^{\mathrm{ele}} / \lambda}]^{T} \otimes [1, \mathrm{e}^{j 2 \pi d \cos \phi_{\mathrm{T}, l}^{\mathrm{ele}} / \lambda}, \! \cdots \! ,  \\
& \mathrm{e}^{j 2 \pi d\left(N_{T2}-1\right) \cos \phi_{\mathrm{T}, l}^{\mathrm{ele}} / \lambda}]^{T}
\end{aligned}
 \end{equation}
where $\lambda$ denotes the wavelength, $d$ denotes the antenna spacing, and $\otimes$ is the Kronecker product.
$\phi_{\mathrm{T},l}^{\mathrm{ele}}$ $(\phi_{\mathrm{T},l}^{\mathrm{azi}})$ denotes the elevation (azimuth) angle-of-departure of the $l$th path for both $\mathbf{h}_{\mathrm{T}, k}$ and $\mathbf{h}_{\mathrm{T}, k_{\text{ele}} }$.
$\boldsymbol{\alpha}_\mathrm{T} ( {{\phi_{\mathrm{T},l}^{\mathrm{azi}}},{\phi_{\mathrm{T},l}^{\mathrm{ele}}}} )$ and $\boldsymbol{\alpha}_\mathrm{R} ( {{\phi_{\mathrm{R},l}^{\mathrm{azi}}},{\phi_{\mathrm{R},l}^{\mathrm{ele}}}} )$  denote the steering vectors at the sender side and the receive side, respectively.

\section{Proposed Channel Estimation Methods}

In this section, we introduce CBDNet, GAN-CBD and MRDN for the cascaded channel estimation of RIS systems.
Leveraging the sparsity of cascaded mmWave channel, we naturally introduce CBDNet-based method into cascaded channel estimation in line with previous works. And we use the GAN structure to improve the network structure.
Specifically, the proposed method MRDN combines the application of residual dense network (RDN) structure and the convolutional block attention module (CBAM) \cite{woo2018cbam}, which serves as a building block and can obtain accurate CSI for the cascaded sparsity channel.
Compared with existing baseline schemes, MRDN can reduce the complexity of RIS hardware.
In the following, we will show the CBDNet, GAN-CBD and MRDN structure channel estimator. In addition, $\mathbf{x}$ and $\mathbf{z}$ represent the input and output of the universal layers and networks, respectively, in this correspondence.

\subsection{CBDNet-based Channel Estimator}

% Fig. \ref{fig:Structure} illustrates the framework of $\textrm{CBDNet}$. To this end,
% We can combine the real and imaginary parts of channel matrix into a larger matrix of size $N_b \times 2 N_u$ first.

$\textrm{DNN}_E$ and $\textrm{DNN}_D$ denote the noise level estimation subnetwork and the non-blind denosing subnetwork, respectively. $\Theta_E$ and $\Theta_D$ are the network parameters of $\textrm{DNN}_E$ and $\textrm{DNN}_D$, respectively.

%Furthermore, $\textrm{DNN}_E$ takes a noisy channel matrix $\mathbf{Y}$ to generate the estimated noise map $\mathbf{M}$ by training $\mathbf{W_E}$. We have $\mathbf{M} = \mathcal{F}_E(\mathbf{Y},\mathbf{W_E})$, where $\mathbf{M} \in \mathbb{R}^{N_b \times 2N_u}$ .
%In order to facilitate the estimation using the full convolution network, the noise level map should be the same size as $\mathbf{Y}$.

%Then, the $\textrm{DNN}_D$ takes both $\mathbf{Y}$ and $\mathbf{M}$ as input to obtain the estimated channel $\mathbf{H} = \mathcal{F_D}(\mathbf{Y},\mathbf{M},\mathbf{W_D})$, where $\mathbf{W_D}$ denotes the network parameters of $\textrm{DNN}_D$.

\subsubsection{Basic Structure}
Assuming that $*$ denotes \emph{Conv} function, as $\mathbf{x}$ and $\mathbf{z}$ are the input and output of the $k$th \emph{Conv} layer, the mathematical deduction for convolutional layer is
\begin{equation}
\begin{aligned}
\mathbf{z} =W_{k} * \mathbf{x}+b_{k},
\end{aligned}
\end{equation}
where the weight and bias matrices $W_{k}$ and $b_{k}$ are the $k$th \emph{Conv} parameters. $\mathbf{z}= c(\mathbf{x})$, $\Theta_{k,c} = \{W_{k,c}, b_{k,c} \}$ for ``\emph{Conv}" layers, $\mathbf{z}= s(\mathbf{x})$, $\Theta_{k,s} = \{W_{k,s}, b_{k,s} \}$ for ``\emph{SoftMax}" layers.
% \Theta_{k} = \{ W_{1,c}, \ldots, W_{k,c}, b_{1,c}, \ldots, b_{k,c} \}$
Assuming that $\max$ denotes ``\emph{ReLU}" layer function, the mathematical deduction for ``\emph{ReLU}" layer is
 \begin{align}
\mathbf{z} =\max \left(0, \mathbf{x}\right),
 \end{align}
count as $\mathbf{z}= r(\mathbf{x})$ for ``\emph{ReLU}" layers.

\subsubsection{Noise Level Estimation Subnetwork}

\begin{itemize}
  \item Input Layer: As the real and imaginary parts of the received signal matrix ${\mathbf{y}_{k}} \in \mathbb{C}^{N_b \times N_u}$ are independent at the BS, we first combine them into a super matrix $\mathbf{Y} \in \mathbb{R}^{N_b \times 2N_u}$ as the input of $\textrm{DNN}_E$. $\mathbb{R}^{{M} \times {N}}$ represents an ${M} \times {N}$ real-valued matrix.
  \item Convolutional sensing:
  The $\textrm{DNN}_E$ consists of $\mathrm{B}_c$ \emph{Conv} layers and $\mathrm{K}$ \emph{SoftMax} layers. The recurrence relation of main body for $\textrm{DNN}_E$ is
\begin{equation}
\begin{aligned}
\sigma&= \mathcal{F}_E(\mathbf{Y},\Theta_E) \\
&=c \! \circ \! \cdots \! \circ c \circ \! s \circ \! \cdots \! \circ s(\mathbf{Y})=(c)^{\mathrm{B}_c}\circ(s)^{\mathrm{K}}(\mathbf{Y}),
\end{aligned}
\end{equation}
where the operator $\circ$ denotes a function composition, $\sigma$ denotes the noise level for the space-invariant AWGN, $\mathbf{M} \in \mathbb{R}^{N_b \times 2N_u}$ is a uniform map where all elements are $\sigma$, $\Theta_E = \{\Theta_{1,c}, \ldots,\Theta_{\mathrm{B}_c,c}, \Theta_{1,s}, \ldots,\Theta_{\mathrm{K},s},\}$. The $\mathcal{F}_E: \mathbb{R}^{N_b \times 2N_u} \mapsto \mathbb{R}^{1 \times 1}$ is the mapping function for $\textrm{DNN}_E$.
\end{itemize}
%Assume that $*$ denotes \emph{Conv} layer function
%$c$ and $s$ denotes a \emph{Conv} and \emph{SoftMax} function, respectively,

\subsubsection{Non-Blind Denosing Subnetwork}

\begin{itemize}
        \item Input Layer:
        The $\textrm{DNN}_D$ takes both $\mathbf{Y}$ and $\mathbf{M}$ as input to obtain the estimated channel $\widehat{\mathbf{H}}$.
        \item Residual Blocks:
        The $\textrm{DNN}_D$ consists of $\mathrm{B}$ residual blocks $c \circ b \circ r $, then, the recurrence relation of main body for $\textrm{DNN}_D$ is
\begin{equation}
\begin{aligned}
\mathbf{{H}}_\mathrm{m}&= \mathcal{F}_D(\mathbf{Y},\mathbf{M},\Theta_D) \\
&=c \circ b \circ r \cdots c \circ b \circ r(\mathbf{Y},\mathbf{M}) \\
&=(c \circ b \circ r)^{\mathrm{B}}(\mathbf{Y},\mathbf{M}).
\end{aligned}
\end{equation}
The middle output $\mathbf{{H}}_\mathrm{m}=\mathcal{F}_D(\mathbf{Y},\mathbf{M},\Theta_D)$, where $\mathcal{F}_D: \mathbb{R}^{N_b \times 2N_u} \mapsto \mathbb{R}^{N_b \times 2N_u}$ is the mapping function for stacking residual blocks.
        \item Output Layer: By reversing the combining, the middle output of $\textrm{DNN}_D$ $ \mathbf{{H}}_\mathrm{m} \in \mathbb{R}^{N_{b} \times 2  N_{u}} $ produces the estimated channel matrix $\mathbf{\widehat{\mathbf{H}}} \in \mathbb{C}^{N_b \times  N_{u}}$.
        \item Loss Function:
        In asymmetric learning, the noise level is estimated to improve the loss function, to quantify the effectiveness of $\textrm{DNN}_D$ criterion. The loss function is denoted as
 \begin{align}
{\mathcal{L}_\text{rec} } = {\frac{1}{\sigma}} \Vert  \mathbf{\widehat{\mathbf{H}}}-\mathbf{H} \Vert ^2
 \end{align}
 Given the estimated noise level $\sigma(\mathbf{Y})$ and the truth $\sigma(\mathbf{Y_i})$, more penalty is incorporated into their $\textrm{MSE}$ when $\sigma(\mathbf{Y}) < \sigma(\mathbf{Y_i})$.
    \end{itemize}

%\textcolor{blue}{$\Theta_D = \{\Theta_{1,c}, \ldots,\Theta_{\mathrm{B},c}\}$} and
\subsection{GAN-based Channel Estimation}
Motivated by the development of generative adversarial networks (GAN) structure technique, based on the previous CBDNet as our own generator subnetwork, we develop our own GAN-CBD for denoise modeling.
%The GAN paradigm estimates generative samplers by means of a training procedure which pits a generator $G$ against a discriminator $D$. $D$ is trained to tell apart training examples from samples produced by $G$, while $G$ is trained to increase the probability of its samples being incorrectly classified as data.
%Likewise with the previous technique, we use residual blocks (i.e., CBDNet) as a building module of the generator. Moreover, we made several modifications to improve the performance of denoise modeling.
The GAN paradigm generates samplers through training and fitting as CBDNet works, and the results of GAN-CBD network compare with and label results, making the discriminator $D$ work well.
Training $D $ can distinguish the training examples from the samples generated by $ G $, and $ G $ undergoes the judgment of $ D $ to reduce the possibility of samples being misclassified.

\subsubsection{Generator Network}
In addition, in order to verify the effectiveness of GAN structure, we use CBDNet as the generator network.
The $\textrm{GAN}_D$ consists of $\mathrm{B}$ residual blocks. We have
\begin{equation}
\begin{aligned}
&\mathbf{\widehat{\mathbf{H}}}= \mathcal{G}_d(\mathbf{Y},\mathbf{M},\Theta_{G_d})\\
&=c \circ b \circ r \cdots c \circ b \circ r(\mathbf{Y},\mathbf{M})=(c \circ b \circ r)^{\mathrm{B}}(\mathbf{Y},\mathbf{M}),
\end{aligned}
\end{equation}
where $\mathcal{G}_d: \mathbb{R}^{N_{b} \times 2  N_{u}} \mapsto \mathbb{C}^{N_{b} \times N_{u}}$ is the mapping function for the generator network.
$\mathbf{M} \in \mathbb{R}^{N_b \times 2N_u}$ is a uniform map from $\textrm{GAN}_E$, $\sigma = \mathcal{G}_e(\mathbf{Y},\Theta_{G_e})$.

\subsubsection{Discriminator Network}
In the original formulation, the training procedure defines a continuous minimax game as
\begin{equation}
\begin{aligned}
\underset{G}{\arg \min}  \ \underset{D}{\arg \max} \mathbb{E} [\log D(\mathbf{x})]+\mathbb{E} [\log (1-D(G(\mathbf{n})))]
\end{aligned}
\end{equation}
where $D$ is a function that maps $\mathbb{R}^{N_b \times 2N_u}$ to the unit interval, and $G$ is a function that maps a noise vector $\mathbf{n} \in \mathbb{R}^{N_b \times 2N_u}$, drawn from a simple distribution $p(\mathbf{n})$, to the ambient space of the training data $\mathbb{R}^{N_b \times 2N_u}$.

\subsection{MRDN-based Channel Estimation}

\begin{figure}[t]
  \centering
\includegraphics[scale=0.43]{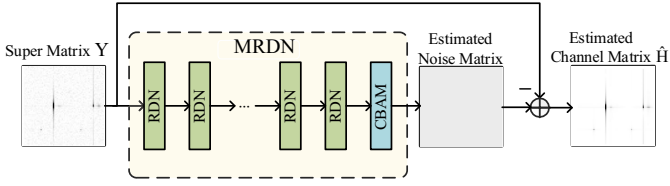}
\caption{MRDN-based channel estimation system.}
\label{fig:MRDN}
\end{figure}

We define this feature concatenation part of RDN and CBAM in Fig. \ref{fig:MRDN}, and use it as a building module of MRDN.

\subsubsection{Basic Structure}

Assuming that $*$ denotes ``\emph{Conv}" layer function, $\max$ denotes ``\emph{ReLU}" layer function, and the residual block model is the composition of two cascaded functions:
 \begin{align}
\mathbf{z}_{-1} &=W_{n, r} * \mathbf{x}+b_{n,r},
 \end{align}
 \begin{align}
\mathbf{z}_{0} &=\max \left(0, \mathbf{z}_{-1}\right),
 \end{align}
where the weight and bias matrices of the $n$th residual block parameter are denoted by $\Theta_{n, r} = \{ W_{n, r}, b_{n,r} \}, n \in \{1,2, \ldots, \mathrm{B} \} $. $\mathbf{x}$ and $\mathbf{z}_{0}$ are the input and output of the residual block, respectively. let $g_n$ denotes the single recursion function of $n$th residual block.

\begin{figure}[t]
  \centering
\includegraphics[scale=0.52]{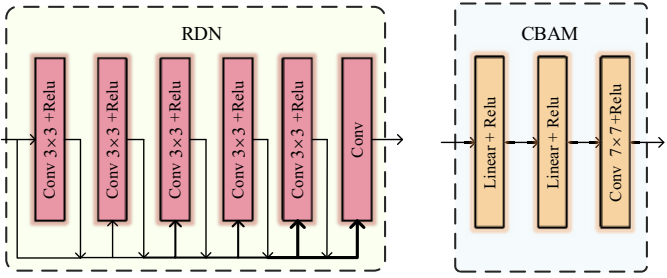}
\caption{RDN system model.}
\label{fig:rdn}
\end{figure}

\subsubsection{Residual Dense Network Structure}
RDN performs well in addressing denoising image problems. Motivated by many recent image restoration networks including RDN, we include the global residual connection such that the network can focus on learning the difference between the noisy and ground-truth channel matrix. The main body of RDN have $\mathrm{B}$ layers. The recurrence relation of main body for the $n$th layer is $F_1 = g_1(\mathbf{Y})$ and
% \begin{align}
%F_1 = g_1(\mathbf{Y}),
% \end{align}
 \begin{align}
F_n = g_n(F_{n-1}(\mathbf{Y}), \cdots, F_{1}(\mathbf{Y}), \mathbf{Y}), \forall n \in \{2, \ldots, \mathrm{B} \}.
 \end{align}

% The RDN consists of $\mathrm{B}$ residual blocks.

\subsubsection{Convolutional Block Attention Module}

 \begin{align}
\mathbf{z}_{-1} &=W_{-1, a} * \mathbf{x}+b_{-1,a},
 \end{align}
 \begin{align}
\mathbf{z}_{0} &=\max \left(0, \mathbf{z}_{-1}\right),
 \end{align}
\begin{align}
\mathbf{z}_{1} &=W_{1, a}  * \mathbf{z}_{0} +b_{1,a},
\end{align}
where the weight and bias matrices consist the CBAM parameter  $\Theta_a = \{ W_{-1, a}, W_{1, a}, b_{-1,a}, b_{1,a} \}$. $\mathbf{x}$ and $\mathbf{z}_{1}$ are the input and output of the CBAM.
The recurrence relation of CBAM for MRDN is $A (\mathbf{x})= c \circ r \circ c(\mathbf{x})$.
% \begin{align}
%K (\mathbf{x})=f \circ r \circ f(\mathbf{x}).
% \end{align}

\begin{figure}[t]
  \centering
  %\fontsize{9}{9}\selectfont
\includegraphics[scale=0.48]{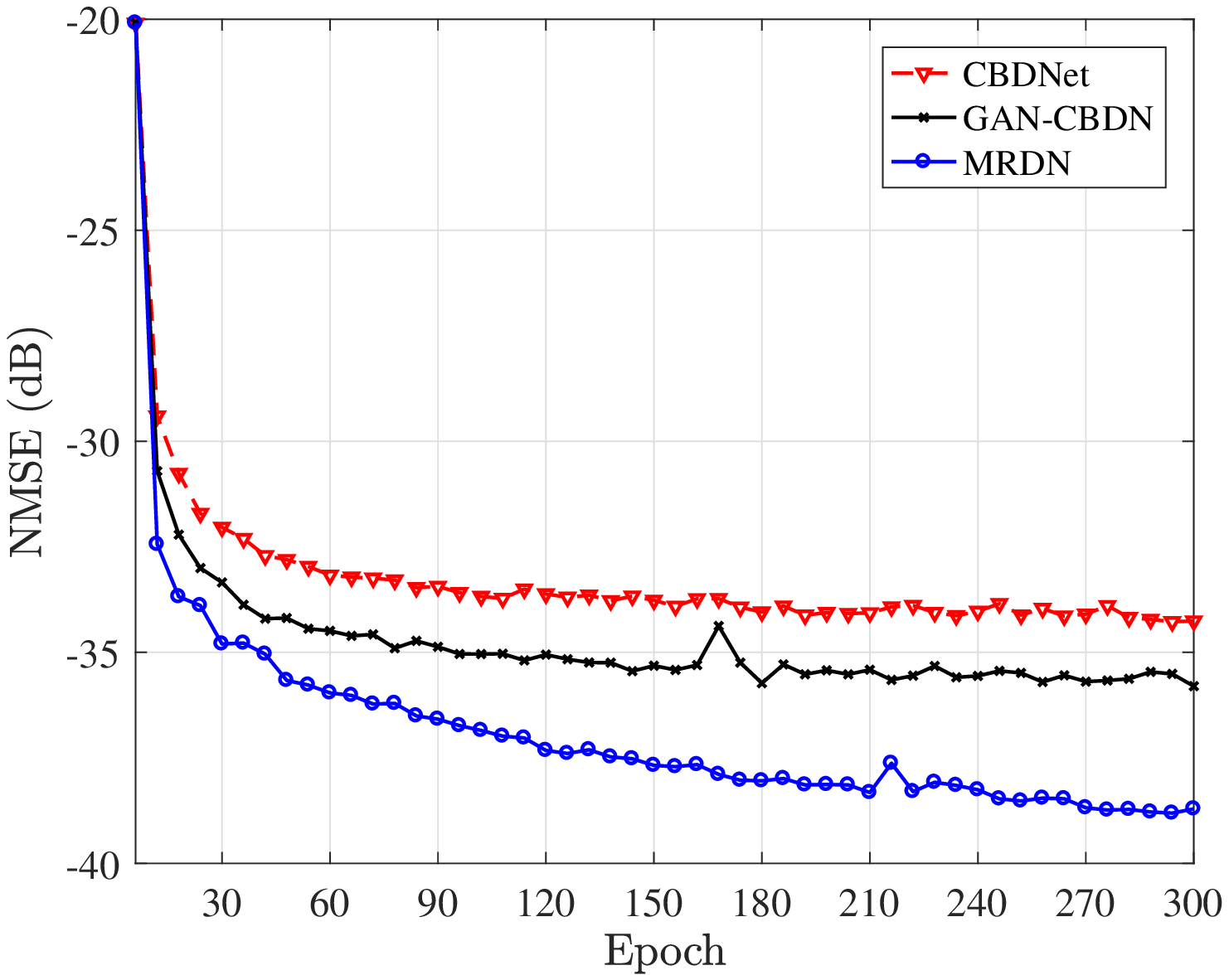}
\caption{Convergence of $\textrm{CBDNet}$, $\textrm{GAN-CBD}$ and $\textrm{MRDN}$.}
\label{fig:mrdN_bubust}
\end{figure}

\begin{figure}[t]
  \centering
  %\fontsize{9}{9}\selectfont
\includegraphics[scale=0.52]{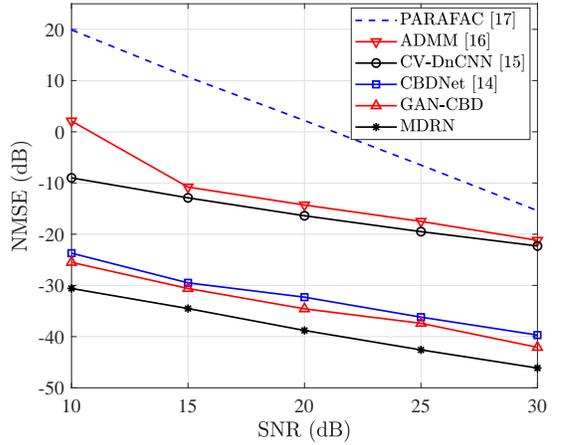}
\caption{NMSE performance comparison of CBDNet, GAN-CBD, MRDN with CS methods.}
\label{fig:method}
\end{figure}

\subsubsection{Input Layer}
As the real and imaginary parts of the received signal matrix ${\mathbf{y}_k} \in \mathbb{C}^{N_{b} \times  N_{u}}$ are independent at the RIS, we first combine them into a super matrix $\mathbf{Y} \in \mathbb{R}^{N_{b} \times 2 N_{u}}$.
In this case, the channel matrix can be treated as a 2D image and the super matrix $ \mathbf{Y} $ is the input of MRDN.

\subsubsection{Multiple Residual Dense Network Structure}
We take advantages of novel ideas in RDN and RCAN as follows.
\begin{itemize}
     \item RDN itself is an image restoration network, but we use it with modifications as a component of our network and construct a cascaded structure of $N_R$ RDNs as our image denoising network.
     \item The recurrence relation of main body for RDN is
 \begin{align}
M(\mathbf{x})= F_{n, N_R} \circ F_{n, N_R-1} \circ \cdots \circ F_{n, 1}(\mathbf{x}),
 \end{align}
 \begin{align}
F(\mathbf{x})=M \circ A(\mathbf{x}) = F_{n}^{N_R} \circ A(\mathbf{x}),
 \end{align}
 where the operator $\circ$ denotes a function composition and $F_{n}^{N_R}$ denotes the $N_R$-fold product of $F_{n}$. The middle output $\mathbf{{H}}_\mathrm{m}=F(\mathbf{Y})$, where $F: \mathbb{R}^{N_{b} \times 2  N_{u}} \mapsto \mathbb{R}^{N_{b} \times 2  N_{u}}$ is the mapping function for MRDN.
\end{itemize}

\subsubsection{Computational Complexity Analysis}
The computational complexity of the training phase in $\textrm {CBDNet}$ is given by
\begin{equation}
\begin{aligned}
 \mathcal{O}(N^2{K^2}s t({L_d} {D_l}^2+{L_e} {E_l}^2)),
\end{aligned}
\end{equation}
where $s$ donates the size of mini-batch, $t$ donates the number of iterations, ${K^2}$ donates the size of kernels. ${L_d}$ and ${L_e}$ denote the number of ``\emph{Conv}" for $\textrm{DNN}_D$ and $\textrm{DNN}_E$, ${D_l}$ and ${E_l}$ denote the number of features for the $l$th layer of $\textrm{DNN}_D$ and $\textrm{DNN}_E$, respectively. The computational complexity of the training phase in $\textrm {GAN-CBD}$ and $\textrm {MRDN}$ are given by
\begin{equation}
\begin{aligned}
   \mathcal{O}(N^2{K^2}s t({L_{g,d}} {D_{g,l}}^2+{L_{g,e}} {E_{g,l}}^2 + {L_{a}} {E_{a}}^2)),
\end{aligned}
\end{equation}
and
\begin{equation}
\begin{aligned}
   \mathcal{O}( N^2{K^2}s t {L_m}^2 {D_m}^2).
\end{aligned}
\end{equation}

\section{Simulation Result}

\begin{figure}[t]
  \centering
\includegraphics[scale=0.52]{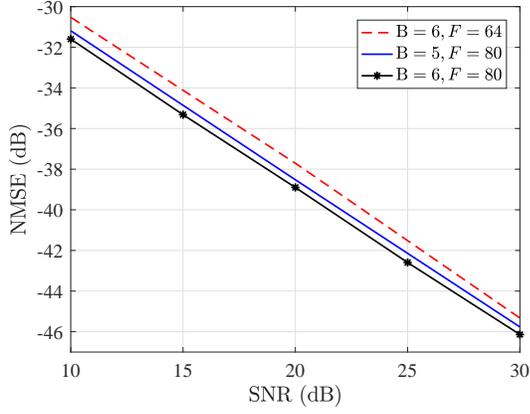}
\caption{NMSE performance of MRDN with different model capacity.}
\label{fig:mrdn_str}
\end{figure}

We consider the RIS-aided mmWave massive $\textrm{MIMO}$ system with 20 UEs, where $N_b = 64$, $N_u = 32$, $N=4096$, $L = 3$ and $d = \lambda / 2$. In terms of hardware, we use Intel Core i7-9700K @3.60GHz, 32 GB RAM and NVIDIA GeForce RTX 2080Ti to implement the above three models through PyTorch library. From the perspective of normalized mean square error (NMSE) performance, this section illustrates the pros and cons of the three proposed channel estimators in terms of structure.
All simulation results are derived in PyCharm Community Edition (Python 3.8 environment). The training rate is set as 0.0001 for MRDN and 0.001 for CBDNet and GAN-CBD and the mini-batch size is 20 for all three methods.
The training, validation, and testing sets include 16,000, 6,000, and 8,000 samples, respectively.
%The training, validation, and testing sets include, respectively.
The training, validation and testing sets for the three methods use the same data set samples.
The number of RDN for MRDN, residual blocks for both CBDNet and GAN-CBD are 6 and 12, respectively. The MRDN has 80 features, CBDNet and GAN-CBD have 96 features.
%In this section, we investigate the NMSE performance of the proposed MRDN over a wide range of SNRs.
Note that $\textrm{NMSE}$ is defined as
%$ \mathbb{E} \left( { {{\left\| {\widehat{ \mathbf{H}} - \mathbf{H}} \right\|^2}}/{{\left\| {\widehat{ \mathbf{H}}} \right\|^2}}} \right)$.
\begin{equation}
   \text{NMSE} = \mathbb{E} \left( { {{\left\| {\widehat{ \mathbf{H}} - \mathbf{H}} \right\|^2}}/{{\left\| {\widehat{ \mathbf{H}}} \right\|^2}}} \right).
\end{equation}

Figure \ref{fig:mrdN_bubust} compares the three different models, including the MRDN, CBDNet and GAN-CBD.
We can find that the MRDN can achieve best NMSE performance and fastest convergence. Because the GAN-CBD brings the advantage of judging the network, it shows better performance than the CBDNet.
The computational complexity of training and offline operation can be hugely reduced. Also, the robustness of the channel estimator to different scenarios is enhanced.
%The average running time (in seconds) of MDSR is 0.0075, respectively.
The average running time of MRDN (in seconds) is 0.0075, while the CBDNet and the GAN-CBD are 0.0094 and 0.0098 respectively, the computational complexity of training and offline operations for the MRDN can reduced compared with the CBDNet and the GAN-CBD.
However, for almost the same computational complexity, the GAN-CBD can achieve better NMSE performance and fast convergence compared with the CBDNet. But compared with the MRDN, the improvement of network structure is not significant.

%Because the GAN structure has a very weak influence on the NMSE performance and the convergence of the neural network compared to b.

Figure \ref{fig:method} compares the $\textrm{NMSE}$ performance of the proposed $\textrm{MRDN}$-based channel estimator for different structures (e.g., $\textrm{CBDNet}$ \cite{jin2019channel}, $\textrm{GAN-CBDN}$, $\textrm{CV-DnCNN}$ \cite{liu2020deep_cvdnn}) and with existing conventional channel estimation methods (e.g., $\textrm{ADMM}$ \cite{vlachos2018massive}, $\textrm{PAPRFAC}$ \cite{wei2020channel}).
The simulation results are averaged over 300 iterations for the three proposed methods.
%It can be observed that MRDN can achieve a best NMSE performance than the CBDNet .
It can be observed that MRDN can achieve better NMSE performance compared with GAN-CBD and CBDNet by 5.63dB and 4.51dB respectively.
Compared with CV-DnCNN, which is also based on CNN, as well as conventional ADMM and PAPRFAC, regardless of the significant performance comparison in NMSE, the lower complexity of MRDN allows it to be better applied.

Figure \ref{fig:mrdn_str} compares the $\textrm{NMSE}$ performance for different number of features and RDN. With more RDNs for the global residual dense connection, and more comprehensive perceptual fields, the MRDN with 80 features and 6 dense connections for RDN performs better.
Consequently, the main challenge in accurately describing noise is the lack of observational dimensions and modeling capabilities of neural networks, such as features and layers.

\section{Conclusion}

We proposed the $\textrm{CBDNet}$, $\textrm{GAN-CBD}$ and $\textrm{MRDN}$ based cascaded channel estimators for RIS-aided mmWave massive MIMO communication systems.
%Utilizing the sparsity of the RIS channel, we transformed the channel estimation problem into a classical problem.
Utilizing the sparsity of the cascaded RIS channels and classic image processing techniques, we regard the channel matrix as a two-dimensional image.
The proposed residual dense network structure can increase the flexibility of the overall network to obtain better generalization and fitting capabilities, while the advantages brought by the GAN structure are not significant.
%Based on this, single-scale and multi-scale deep super-resolution neural networks were designed to estimate the RIS channel. Simulation results verified that the performance of the proposed estimators improves with the density of the sensing devices and the scale of residual dense neural networks.
Compared with the previous generation method, based on the above advantages, the MRDN-based deep learning network is designed to estimate the cascaded RIS channels.
%The simulation results show that the performance of the proposed estimator increases with the density of sensing devices and the scale of the residual dense neural network.
The simulation results show that the performance of the proposed the MRDN estimator increases with the increase of the scale of the network structure under the same order of complexity as the CBDNet and the GAN-CBD.

\begin{appendices}
\section{Forward and Backward Propagation}
Assume that the weight matrix $W_{i} \in \mathbb{R}^{n_{i+1} \times n_{i}}$ and the bias vector $b_{i} \in \mathbb{R}^{n_{i+1}}$ are the parameters at $i$ ``\emph{Conv}" layer $c_{i}$.
In the multilayer perceptron (MLP), we can explicitly write
$c_{i}\left(\mathbf{x}_{i} ; W_{i}, b_{i}\right)=r_{i}\left(W_{i} \cdot \mathbf{x}_{i}+b_{i}\right)$, where $r_{i}$ is an elementwise function and the definition for \emph{Conv} is $r _{i}(\mathbf{x}) \equiv \max (0, \mathbf{x})$, as the first derivative is $r_{i}^{\prime}(\mathbf{x})=H(\mathbf{x})$. For any $\mathbf{x}_{i} \in \mathbb{R}^{n_{i}}$ and vectors $U_{i} \in \mathbb{R}^{n_{i+1} \times n_{i} }$ in inner product space,
\begin{equation}
\begin{aligned}
\nabla_{W_{i}} c_{i}\left(\mathbf{x}_{i}\right) \cdot U_{i}=\mathrm{D} \mit\Psi_{i}\left(\mathbf{z}_{i}\right) \cdot U_{i} \cdot \mathbf{x}_{i},
\end{aligned}
\end{equation}
\begin{equation}
\begin{aligned}
\nabla_{b_{i}} c_{i}\left(\mathbf{x}_{i}\right)=\mathrm{D} \mit\Psi_{i}\left(\mathbf{z}_{i}\right),
\end{aligned}
\end{equation}
where $\mathbf{z}_{i}=W_{i} \cdot \mathbf{x}_{i}+b_{i}$, $\mit\Psi(v)=\sum_{k=1}^{n} \psi\left(v_{k}\right) e_{k}$ and
\begin{equation}
\begin{aligned}
\mathrm{D} f(x ; \theta) \cdot v=\left.\frac{\mathrm{d}}{\mathrm{d} t} f(x+t v ; \theta)\right|_{t=0}.
\end{aligned}
\end{equation}
The loss function gradients in MLP is
\begin{equation}
\begin{aligned}
J(\mathbf{x}, \!\mathbf{z} ; \!\theta) \!=\!\frac{1}{2}\|\mathbf{z} \!\!-\!\!F(\mathbf{x} ; \!\theta)\|^{2}\!=\!\frac{1}{2}\langle \mathbf{z} \!\!-\!\!F(\mathbf{x} ; \!\theta), \mathbf{z} \!\!-\!\!F(\mathbf{x} ; \!\theta)\rangle.
\end{aligned}
\end{equation}
Let $(\mathbf{x}, \mathbf{z}) \in E_{1} \times E_{L+1}$ be a network input-output pair,
\begin{equation}
\begin{aligned}
\nabla_{W_{i}} J(\mathbf{x}, \mathbf{z} ; \theta)=\left[\mit\Psi_{i}^{\prime}\left(\mathbf{z}_{i}\right) \odot\left(\mathrm{D}^{*} \omega_{i+1}\left(\mathbf{x}_{i+1}\right) \cdot e\right)\right] \mathbf{x}_{i}^{T},
\end{aligned}
\end{equation}
\begin{equation}
\begin{aligned}
\nabla_{b_{i}} J(\mathbf{x}, \mathbf{z} ; \theta)=\mit\Psi_{i}^{\prime}\left(\mathbf{z}_{i}\right) \odot\left(\mathrm{D}^{*} \omega_{i+1}\left(\mathbf{x}_{i+1}\right) \cdot e\right),
\end{aligned}
\end{equation}
where $\mathbf{x}_{i}\!=\!\alpha_{i-1}(\mathbf{x})$, and the prediction error is $e \!=\! F(\mathbf{x} ; \! \theta)-\mathbf{z}$,
\begin{equation}
\begin{aligned}
F(\mathbf{x} ; \theta)=\left(c_{L} \circ \cdots \circ c_{1}\right)(\mathbf{x}),
\end{aligned}
\end{equation}
for all $i \in [L]$ and $\theta \in \{W_i,b_i\}$,
\begin{equation}
\begin{aligned}
\nabla_{\theta_{i}} J(\mathbf{x}, \mathbf{z} ; \theta)=\nabla_{\theta_{i}}^{*} c_{i}\left(\mathbf{x}_{i}\right) \cdot \mathrm{D}^{*} \omega_{i+1}\left(\mathbf{x}_{i+1}\right) \cdot e.
\end{aligned}
\end{equation}
The generic layer of a CNN as a parameter-dependent map that takes as input an $m_1$-channeled tensor, where each channel is a matrix of
size $n_{1} \times \ell_{1}$, and outputs an $m_2$-channeled tensor, where each channel is a matrix of size $n_{2} \times \ell_{2}$.
The parameters $W \in \mathbb{R}^{p \times q} \otimes \mathbb{R}^{m_{2}}$, the input $x \in \mathbb{R}^{n_{1} \times \ell_{1}} \otimes \mathbb{R}^{m_{1}}$. $\left\{e_{j}\right\}_{j=1}^{m_{1}}$ denotes an orthonormal basis for
$\mathbb{R}^{m_{1}}$, and $\left\{\bar{e}_{j}\right\}_{j=1}^{m_{2}}$ denotes an orthonormal basis for $\mathbb{R}^{m_{2}}$, the $\mathbf{x}$ and $W$ is:
\begin{equation}
\begin{aligned}
\mathbf{x}=\sum_{j=1}^{m_{1}} \mathbf{x}_{j} \otimes e_{j}, \quad W=\sum_{j=1}^{m_{2}} W_{j} \otimes \bar{e}_{j}.
\end{aligned}
\end{equation}
And the convolution operator $C$ can be written as
\begin{equation}
\begin{aligned}
C(W, \mathbf{x})=\sum_{j=1}^{m_{2}} c_{j}(W, \mathbf{x}) \otimes \bar{e}_{j},
\end{aligned}
\end{equation}
where $c_{j}$ is a bilinear operator that defines the mechanics of the convolution:
\begin{equation}
\begin{aligned}
c_{j}(W, \mathbf{x})=\sum_{k=1}^{\widehat{n}_{1}} \sum_{l=1}^{\widehat{\ell}_{1}}\left\langle W_{j}, \mathcal{K}_{\gamma(k, l, \Delta)}(\mathbf{x})\right\rangle \widehat{E}_{k, l},
\end{aligned}
\end{equation}
where $\mathcal{K}$ is the cropping operator that defines the action of convolution, $\gamma(k, l, \Delta)=(1+(k-1) \Delta, 1+(l-1) \Delta)$, $\Delta$ defines the stride of the \emph{Conv}.
The generic layer $c_{i}$ is
\begin{equation}
\begin{aligned}
c_{i}\left(\mathbf{x}_{i}\right)=\mit\Psi_{i}\left(C_{i}\left(W_{i}, \mathbf{x}_{i}\right)\right).
\end{aligned}
\end{equation}
For the ``\emph{Conv}" layer,
\begin{equation}
\begin{aligned}
\nabla_{W_{i}}^{*} c_{i}(x_{i}) = (\mit\Psi_{i}\llcorner \mathbf{x}_{i})^{*}  \mathrm{D} C_{i}(W_{i}, \mathbf{x}_{i}) \mathrm{D}^{*} \mit\Psi_{i}(C_{i}(W_{i}, \mathbf{x}_{i})).
\end{aligned}
\end{equation}
\begin{equation}
\begin{aligned}
\!\!\!\mathrm{D}^{*} c_{i}(\mathbf{x}_{i}) \!= \!(W_{i}\lrcorner C_{i})^{\!*}  \mathrm{D} \mit\Psi_{i}(C_{i}(W_{i}, \!\mathbf{x}_{i})) \mathrm{D}^{*} \mit\Psi_{i}(C_{i}(W_{i},\! \mathbf{x}_{i})).
\end{aligned}
\end{equation}
where $(e_{1}\lrcorner B) \cdot e_{2}=B(e_{1}, e_{2})$, and $(B\llcorner e_{2}) \cdot e_{1}=B(e_{1}, e_{2})$. The learning rate $\eta \in \mathbb{R}_{+}$, the gradient descent step algorithm update the parameter for backpropagation is
\begin{equation}
\begin{aligned}
\nabla_{W_{i}} J(\mathbf{x}, \mathbf{z} ; \theta) \leftarrow\!(C_{i}\llcorner \mathbf{x}_{i})^{*} \mathrm{D} C_{i}\left(W_{i}, \mathbf{x}_{i}) \mathrm{D}^{*} \mit\Psi_{i}(\mathbf{z}_{i}\right) e_{i},
\end{aligned}
\end{equation}
the parameters can be update by $W_{i} \leftarrow W_{i}-\eta \nabla_{W_{i}} J(x, y ; \theta)$.
Due to the application of the derivative chain rule and error backpropagation, the high-dimensional neural network demonstrates excellent results.
\end{appendices}

\bibliographystyle{IEEEtran}
\bibliography{IEEEabrv,Ref}
\end{document}